\documentclass{elsart}

\usepackage{graphicx}

\usepackage{amssymb}

\begin{document}

\begin{frontmatter}



\title{Testing a hadronic rescattering model for RHIC collisions
using the subdivision method}


\author{T. J. Humanic}
\address{Department of Physics,
The Ohio State University, Columbus, OH 43210
\ead{humanic@mps.ohio-state.edu}\thanksref{nsf}\thanksref{dm}}
\thanks[nsf]{This work was supported by the U. S. National Science
Foundation under grant PHY-0099476.}
\thanks[dm]{The author wishes to thank D\'enes Moln\'ar for helpful
discussions on the subdivision method.}

\begin{abstract}
Recently it has been shown that calculations based on a hadronic 
rescattering model agree rather well with experimental results
from the first RHIC run. Because of the large 
particle densities intrinsically present at the early time steps of 
Monte Carlo calculations attempting to model RHIC collisions
undesirable artifacts resulting in non-causality may be present. 
These effects may compromise the results of
such calculations. The subdivision
method, which can remove such artifacts, has been used to test the
present rescattering model calculations. It is shown that no appreciable
changes are seen in the present calculations in using the subdivision
method, thus strengthing the confidence in the results
of this rescattering model for RHIC.
\end{abstract}

\end{frontmatter}

\section{Introduction}
Recent hadronic rescattering model calculations have been shown to agree
reasonably well with experimental results obtained in the first 
physics run for the Relativistic Heavy Ion Collider (RHIC) 
with $\sqrt{s}$ = 130 GeV Au + Au 
collisions \cite{Humanic:2002b}\cite{Humanic:2002a}.
More specifically, these calculations provide good representations 
of the data for 1) the particle mass dependence of the 
$p_T$ distribution slope parameters (i.e. radial flow),
2) the $p_T$, particle mass, and pseudorapidity dependences of the 
elliptic flow, and 3) the $p_T$ and centrality dependences of the
pion Hanbury-Brown-Twiss (HBT) measurements. 
This is a remarkable accomplishment for a single model.

Many hadrons are initially produced in a relatively small
volume in RHIC-type collisions, particularly at early times in the
interaction. It is a challenging task for a Monte
Carlo calculation of this type to deal accurately with 
binary collisions between particles in such
a large particle density environment without introducing numerical 
artifacts which
may affect the results. One such undesirable artifact that can occur
when the interaction range between two particles is much greater
than the scattering mean-free-path is non-causality of the collisions
\cite{Zhang:1998a}\cite{Molnar:2000a}. The method of subdivision
can be used to minimize these artifacts associated with high particle
density\cite{Zhang:1998b}.

The goal of this work is to apply the method of subdivision to
the present RHIC-energy rescattering calculations to test whether
the results from these calculations without subdivision are valid.
Calculations will be compared with and without subdivision for
the pion observables 1) $(1/{p_T})dN/d{p_T}$, 2) elliptic flow vs.
$p_T$ and $\eta$, and 3) HBT vs. $p_T$, at a collision impact 
parameter of 8 fm. Section II will describe the
calculational methods used and Section III with give 
the results of the study.

\section{Calculational Methods}
\subsection{Hadronic rescattering calculation}
A brief description of the rescattering model
calculational method is given
below. The method used is similar to that used in previous 
calculations for lower CERN Super Proton Synchrotron (SPS)
energies \cite{Humanic:1998a}. 
Rescattering is simulated with a semi-classical 
Monte Carlo calculation which assumes strong binary collisions 
between hadrons. The Monte Carlo calculation is
carried out in three stages: 1) initialization and hadronization, 2)
rescattering and freeze out, and 3) calculation of experimental 
observables. Relativistic kinematics is used 
throughout.  All calculations are made to simulate RHIC-energy
Au+Au collisions in order to compare with the results of the
$\sqrt{s}$ = 130 GeV RHIC data.

The hadronization model employs simple parameterizations to describe the 
initial momenta and space-time of the hadrons similar to
that used by Herrmann and Bertsch \cite{Herrmann:1995a}. The initial 
momenta are assumed to follow a thermal transverse
(perpendicular to the beam direction)
momentum distribution for all particles,
\begin{equation}
(1/{m_T})dN/d{m_T}=C{m_T}/[\exp{({m_T}/T)} \pm 1]
\end{equation}
where ${m_T}=\sqrt{{p_T}^2 + {m_0}^2}$ is the transverse mass, $p_T$ 
is the transverse momentum, $m_0$ is the particle rest mass, $C$ is 
a normalization constant, and $T$ is the initial 
``temperature parameter''
of the system, 
and a gaussian rapidity distribution for mesons,
\begin{equation}
dN/dy=D \exp{[-{(y-y_0)}^2/(2{\sigma_y}^2)]}
\end{equation}
where $y=0.5\ln{[(E+p_z)/(E-p_z)]}$ is the rapidity, $E$ is the 
particle energy, 
$p_z$ is the longitudinal (along the beam direction)
momentum, $D$ is a normalization constant, 
$y_0$ is the central
rapidity value (mid-rapidity), and $\sigma_y$ is the rapidity width.
Two rapidity distributions for baryons have been tried: 1) flat
and then falling off near beam rapidity and 2) peaked at central
rapidity and falling off until beam rapidity. Both baryon
distributions give about the same results. 
The initial space-time of the
hadrons for $b=0$ fm (i.e. zero impact parameter or central collisions) 
is parameterized as having cylindrical symmetry with 
respect to the 
beam axis. The transverse particle density dependence is assumed 
to be that of a
projected uniform sphere of radius equal to the projectile radius, $R$ 
($R={r_0}A^{1/3}$, where ${r_0}=1.12$ fm and $A$ is the
atomic mass number 
of the projectile). For $b>0$ (non-central collisions) the transverse
particle density is that of overlapping projected spheres whose
centers are separated by a distance b. The particle multiplicities
for $b>0$ are scaled from the $b=0$ values by the ratio of the
overlap volume to the volume of the projectile.
The longitudinal
particle hadronization position ($z_{had}$) and time ($t_{had}$) 
are determined by the relativistic equations \cite{Bjorken:1983a},
\begin{equation}
z_{had}=\tau_{had}\sinh{y};   t_{had}=\tau_{had}\cosh{y}
\end{equation}
where $y$ is the particle rapidity and $\tau_{had}$ is the 
hadronization proper time.
Thus, apart from particle multiplicities, the hadronization model has 
three free
parameters to extract from experiment: $\sigma_y$,
$T$ and $\tau_{had}$.
The hadrons included in the calculation are pions, kaons,
nucleons and lambdas
($\pi$, K, N, and $\Lambda$), and the $\rho$, $\omega$, $\eta$, 
${\eta}'$, 
$\phi$, $\Delta$, and $K^*$ resonances. For simplicity, the
calculation is isospin averaged (e.g. no distinction is made among
a $\pi^{+}$, $\pi^0$, and $\pi^{-}$). Resonances are present at 
hadronization
and also can be produced as a result of rescattering. Initial resonance
multiplicity fractions are taken from 
Herrmann and Bertsch \cite{Herrmann:1995a}, 
who extracted results from the HELIOS experiment \cite{Goerlach:1992a}. 
The initial resonance fractions used in
the present calculations are: $\eta/\pi=0.05$, $\rho/\pi=0.1$, 
$\rho/\omega=3$, $\phi/(\rho+\omega)=0.12$,
${\eta}'/\eta=K^*/\omega=1$ and, for simplicity, $\Delta/N=0$.

The second stage in the calculation is rescattering 
which finishes with the
freeze out and decay of all particles. Starting 
from the initial stage ($t=0$ fm/c), the positions 
of all particles are allowed to evolve in time in small
time steps ($dt=0.1$ fm/c) according to their 
initial momenta. At each time step
each particle is checked to see a) if it decays, and b) if it is 
sufficiently
close to another particle to scatter with it.
Isospin-averaged s-wave and p-wave cross sections 
for meson scattering are 
obtained from Prakash et al.
\cite{Prakash:1993a}. The calculation is carried out to 100 fm/c,
although most of the rescattering finishes by about 30 fm/c.
The rescattering calculation
is described in more detail elsewhere \cite{Humanic:1998a}.

Calculations are carried out assuming initial parameter values and particle
multiplicities for each type of particle. In the last stage of the 
calculation, the freeze-out and decay momenta and 
space-times are used to produce
observables such as pion, kaon, and nucleon
multiplicities and transverse momentum and rapidity distributions. 
The values of the 
initial parameters of the calculation and multiplicities
are constrained to give observables which agree with 
available measured hadronic observables. As a cross-check
on this, the total kinetic energy from the calculation is 
determined and
compared with the RHIC center of mass energy of 
$\sqrt{s}=130$ GeV to see that they are in 
reasonable agreement. Particle multiplicities were estimated from
the charged hadron multiplicity measurements of the RHIC 
PHOBOS experiment \cite{Back:2000a}. Calculations were carried
out using isospin-summed events containing at
freezeout for central collisions ($b=0$ fm) about 5000 pions, 
500 kaons, and 650 nucleons
($\Lambda$'s were decayed).
The hadronization model parameters used were $T=300$ MeV,
$\sigma_y$=2.4, and $\tau_{had}$=1 fm/c. It is interesing
to note that the same value of $\tau_{had}$ was required 
in a previous rescattering calculation to successfully 
describe results from SPS
Pb+Pb collisions \cite{Humanic:1998a}. 

\subsection{Subdivision method}
The method of subdivision is based on the invariance of
Monte Carlo particle-scattering calculations for a simultaneous 
decrease of the
scattering cross sections by some factor, $l$, and increase of the
particle density by $l$\cite{Zhang:1998a}. As $l$ becomes sufficiently
large, non-causal artifacts become insignificant. The present
rescattering calculation will be tested comparing pion observables
for no-subdivision, i.e. $l=1$, with subdivision of $l=5$. A subdivision
of $l=5$ was chosen since when applied
to other Monte Carlo scattering calculations this value has been shown 
to produce large changes in observables when non-causal artifacts 
are present\cite{Zhang:1998a}\cite{Molnar:2000a}. Since the particle
density increase is accomplished by increasing the particle number 
by a factor $l$, the computer CPU time taken per event increases by
a factor $l^2$, effectively ``slowing down'' a
subdivision study by a factor $l$ compared with not using subdivision.
For the present study, 1640 events and 124 events were generated for 
the $l=1$ and $l=5$ samples, respectively (the statistical value of
the $l=5$ events being 5 times greater per event). The CPU time taken
to generate the $l=5$ sample of events was 420 CPU-hours on a
1 GHz PC processor.

\section{Results}
Figures 1-4 show comparisons of $l=1$ with $l=5$ hadron rescattering
model calculations for pion observables. Descriptions of how the
observables are extracted from the rescattering calculation
are given elsewhere\cite{Humanic:2002a}. All calculations are carried
out for an impact parameter of 8 fm to simulate a medium non-central 
collision which should result in significant elliptic flow for the 
purposes of the present test. Statistical errors are shown either as
error bars or are of the order of the marker size when error bars are
not shown.

Figure 1 compares the $p_T$ distributions for a rapidity cut
of $-2<y<2$. As seen, the $l=1$
and $l=5$ calculations agree within statistical errors, although
there is a slight
decreasing trend for $l=5$ compared with $l=1$ for 
$p_T > 3$ GeV/c.

Figures 2 and 3 compare elliptic flow ($V_2$). Figure 2 shows
$V_2$ vs. $p_T$ for a rapidity cut of $-2<y<2$ and Figure 3 
shows $V_2$ vs. $\eta$ for a $p_T$ cut of $0<p_T<3$ GeV/c. For both
cases, the $l=1$ and $l=5$ calculations are seen to agree within 
statistical errors, reproducing the ``flattening'' in $V_2$ at large
$p_T$ and the ``peaked'' dependence of $V_2$ on $\eta$.

Finally, Figure 4 compares HBT results vs. $p_T$ for a rapidity
cut of $-2<y<2$. Each result is calculated in a $p_T$ bin of 0.2 GeV/c.
A three-dimensional fit is made to extract the pion source parameters
$R_{Tside}$, $R_{Tout}$, $R_{long}$, and $\lambda$. As seen in this
figure, the $l=1$ and $l=5$ calculations agree within
statistical errors for all parameters at all $p_T$ bins, reproducing
the decreasing trend of the $R$-parameters
and increasing trend of the $\lambda$-parameter with increasing $p_T$.

It can be concluded from Figures 1-4 that non-causal artifacts
of the type which can be removed by the subdivision method are
not present in the hadronic rescattering code used in this study.
It is possible to speculate on why this is the case. There are three
main features of the code which might tend to reduce these artifacts:
1) individual particles are allowed to scatter only once per time step,
2) a ``scattering time'' of two time steps is defined during
which particles that have scattered are not allowed to rescatter,
and 3) once two particles have scattered with each other, they are 
not allowed to scatter with each other again in the calculation.
The present study was performed for non-central collisions 
($b=8$ fm), but it is expected that the same conclusion, i.e. that
subdivision does not significantly effect the results of the calculation,
would also be obtained for central collisions. This is because
the particle density is not very different for central
collisions in the model than for mid-peripheral collisions since
the particle multiplicities are scaled by the overlap volume 
for $b>0$, as mentioned earlier.

In summary, the subdivision method has been used to test the validity 
of the present rescattering model calculations. It is found that
no appreciable
changes are seen comparing $l=1$ and $l=5$ calculations, 
thus strengthing the confidence in the results presented previously
from this rescattering model for RHIC.

\begin{figure}
\begin{center}
\includegraphics[width=140mm]{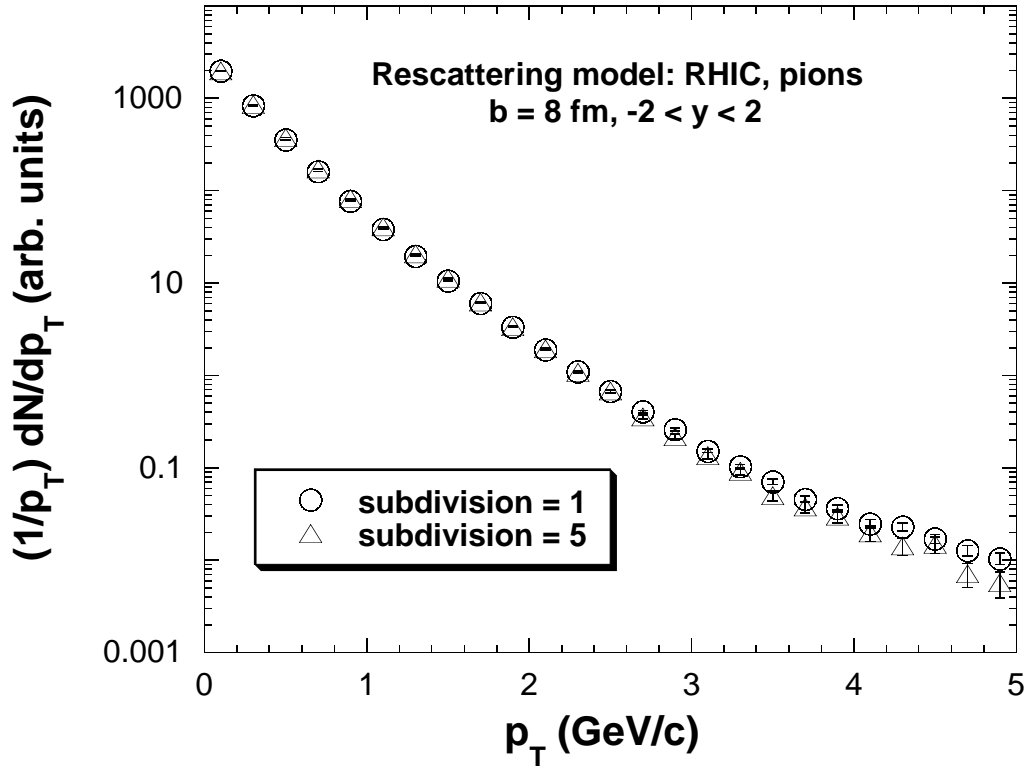}
\caption{\label{fig:sd1} Pion $p_T$ distributions for
$l=1$ and $l=5$.}
\end{center}
\end{figure}

\begin{figure}
\begin{center}
\includegraphics[width=140mm]{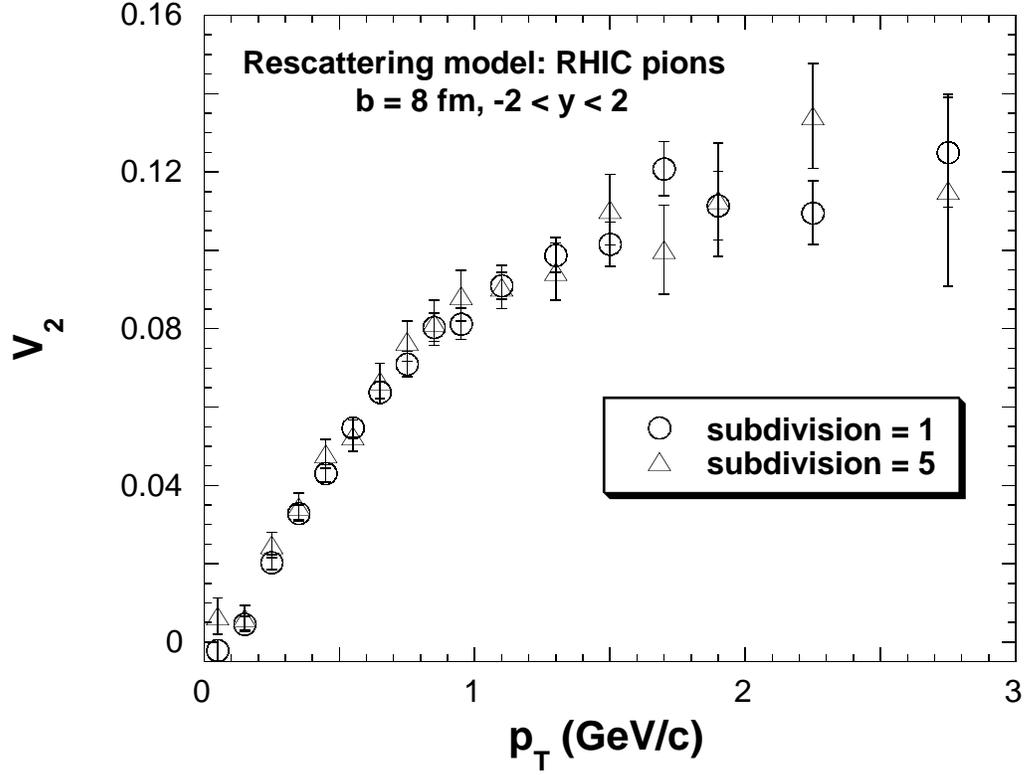}
\caption{\label{fig:sd2} Pion elliptic flow vs. $p_T$
for $l=1$ and $l=5$.}
\end{center}
\end{figure}

\begin{figure}
\begin{center}
\includegraphics[width=140mm]{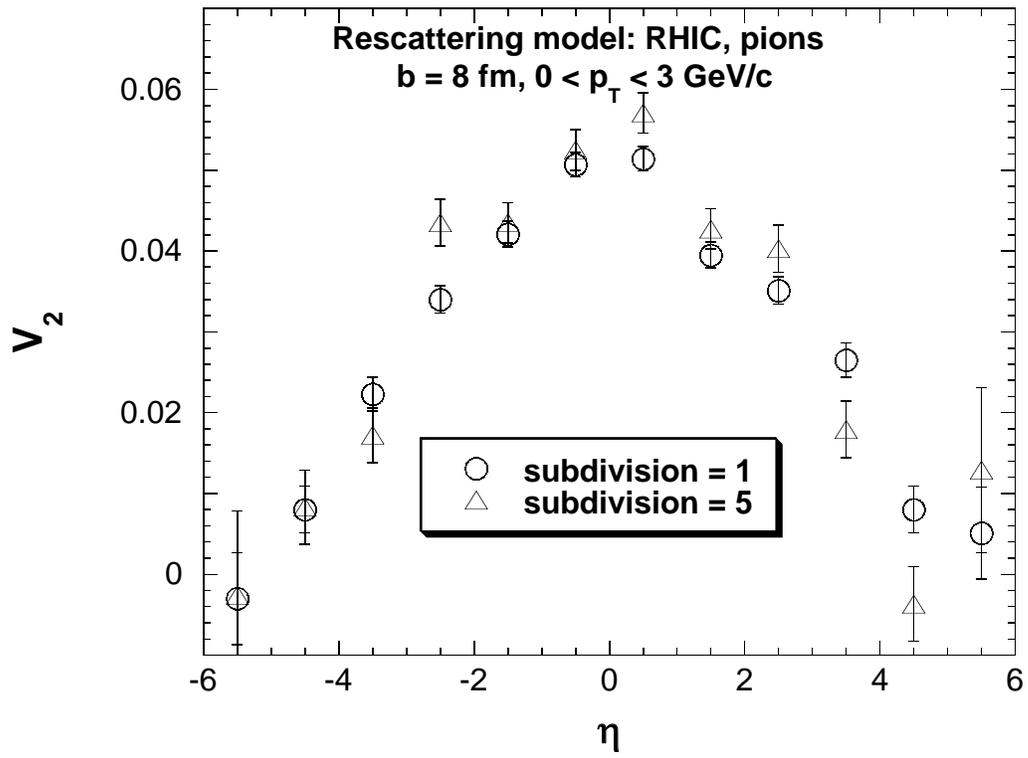}
\caption{\label{fig:sd3} Pion Elliptic flow vs. pseudorapidity ($\eta$)
for $l=1$ and $l=5$.}
\end{center}
\end{figure}

\begin{figure}
\begin{center}
\includegraphics[width=140mm]{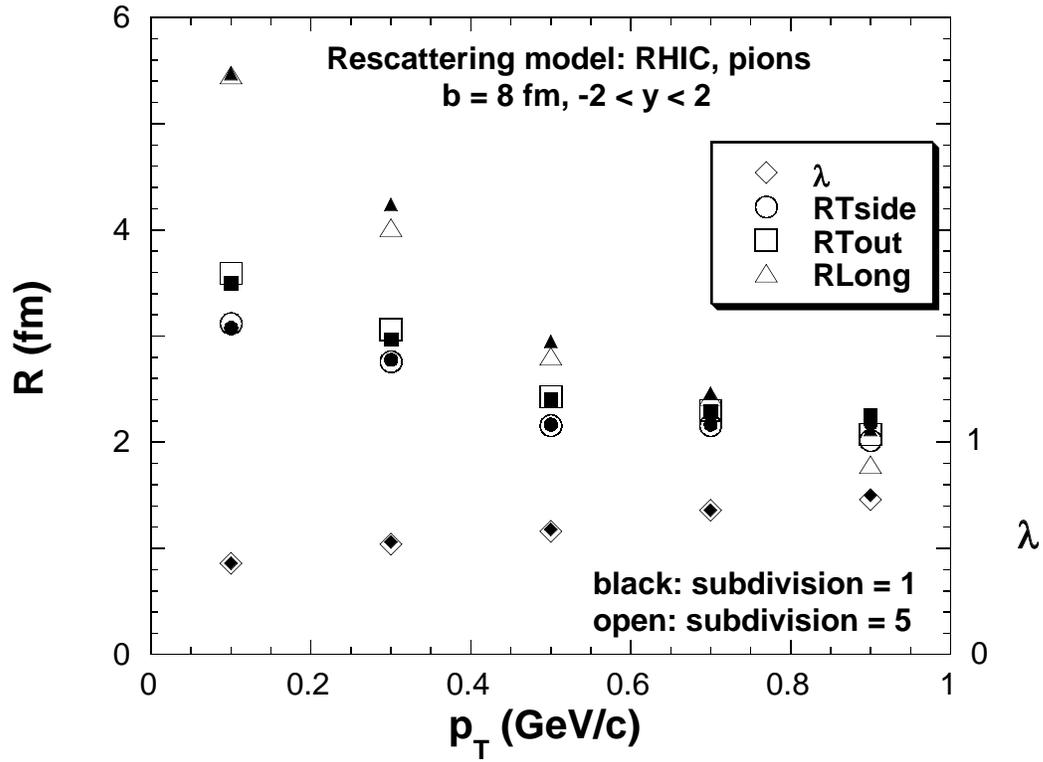}
\caption{\label{fig:sd4} Pion HBT vs. $p_T$ for $l=1$ and $l=5$.
The ordinate scale for $\lambda$ is shown to the right.}
\end{center}
\end{figure}


\end{document}